\documentclass[grl]{agutex2015}

 \usepackage{lineno}
  \usepackage[dvips]{graphicx}
 \setkeys{Gin}{draft=false}
\authorrunninghead{LE ET AL.}

\titlerunninghead{MIXING AND HEATING IN 3D RECONNECTION}

\begin{document}

%
%

\title{Enhanced Electron Mixing and Heating in 3D Asymmetric Reconnection at the Earth's Magnetopause}
%
%

%
%



\authors{A. Le\altaffilmark{1},
 W. Daughton\altaffilmark{1}, L.-J. Chen\altaffilmark{2}, and J. Egedal\altaffilmark{3}}

\altaffiltext{1}{Plasma Theory and Applications, Los Alamos National Laboratory,
Los Alamos, New Mexico 87545, USA.}

\altaffiltext{2}{Astronomy Department, University of Maryland, College Park, MD 20742, USA and NASA Goddard Space Flight Center, Greenbelt, MD 20771, USA.}

\altaffiltext{3}{Department of Physics, University of Wisconsin--Madison,
Madison, Wisconsin 53706, USA.}

%
%


 \keypoints{\item A fully kinetic 3D simulation resolves electron mixing and heating at a magnetopause reconnection site for parameters matching MMS observations. 
 \item Lower-hybrid range drift turbulence mixes plasma across the magnetospheric separatrix.
 \item Strong parallel electron heating near the current layer, as observed by MMS, is reproduced only in 3D.
}


%
%


\begin{abstract}
Electron heating and mixing during asymmetric reconnection are studied with a 3D kinetic simulation that matches plasma parameters from Magnetospheric Multiscale (MMS) spacecraft observations of a magnetopause diffusion region. The mixing and heating are strongly enhanced across the magnetospheric separatrix compared to a 2D simulation. The transport of particles across the separatrix in 3D is attributed to lower-hybrid drift turbulence excited at the steep density gradient near the magnetopause. In the 3D simulation (and not the 2D simulation),  the electron temperature parallel to the magnetic field within the mixing layer is significantly higher than its upstream value in agreement with the MMS observations.
\end{abstract}

%
%

%

\begin{article}

%
%

\section{Introduction}

Magnetic reconnection transports plasma and energy from the shocked solar wind of the magnetosheath across the magnetopause into Earth's magnetosphere \citep{paschmann:1979}. The Magnetospheric Multiscale (MMS) mission \citep{burch:2014} has enabled detailed multi-point observations of this process with instruments that resolve plasma kinetic scales. The first phase of the mission focused on the magnetopause, which is characterized by large asymmetries in conditions between the sheath and magnetospheric plasmas. Particle-in-cell simulation has been used to study the kinetic signatures of asymmetric reconnection \citep{pritchett:2008,mozer:2009grl,malakit:2010,egedal:2011pop,pritchett:2012,hesse:2014,shay:2016,chen:2016b}, primarily in two dimensional (2D) systems, and many key particle and field signatures agree favorably with observations \citep{chen:2016,egedal:2016}.

An important class of fluctuations is suppressed in 2D, however. Drift waves and instabilities, particularly the lower-hybrid drift instability (LHDI) \citep{krall:1971}, entail variations out of the 2D plane, and may interact with reconnecting current sheets \citep{huba:1977,hoshino:1991,lapenta:2003,daughton:2004}. There have been numerous observations of lower-hybrid fluctuations at magnetospheric reconnection sites \citep{gary:1979,andre:2001,bale:2002,vaivads:2004,zhou:2009,norgren:2012,graham:2016,graham:2017} and in laboratory reconnection experiments \citep{carter:2001,fox:2010}. Initial studies focused on the possibility that LHDI fluctuations could contribute to anomalous resistivity \citep{davidson:1975} and thereby help explain fast reconnection rates. Under typical magnetospheric conditions with $T_i/T_e\ge1$, however, LHDI does not significantly alter the reconnection rate or gross dynamics of reconnecting current sheets \citep{roytershteyn:2012,zeiler:2002,pritchett:2013}.  

We re-consider here the role of drift fluctuations using fully kinetic simulations that match the upstream plasma conditions of a reconnection event observed by the MMS spacecraft \citep{burch:2016,price:2016}. For this recent MMS event, simulations indicate that 3D instabilities do indeed contribute to average electron momentum balance through so-called anomalous viscosity \citep{price:2016}, although the overall reconnection rate is nevertheless similar to 2D. The lower-hybrid range fluctuations are found to enhance the transport of plasma across the magnetospheric separatrix. The plasma transport is then linked with parallel (to the local magnetic field) heating of the electrons near the magnetopause, as observed by MMS. In the 2D simulation, the observed electron heating scales with previously derived equations of state for adiabatically trapped electrons \citep{le:2009,egedal:2013pop} on both sides of the current sheet. In 3D, lower-hybrid fluctuations with fast parallel dynamics cause the magnetosphere-side electron temperature to follow Chew-Goldberger-Low (CGL) scalings \citep{chew:1956}.

\section{Particle-in-Cell Simulations}

Asymmetric reconnection is studied with fully kinetic simulations in 2D and 3D geometries using the code VPIC \citep{bowers:2008}. The upstream plasma conditions are very similar to those considered in previous numerical modeling \citep{price:2016} of the MMS diffusion region encounter 16 October 2015 reported by \citet{burch:2016}. The density asymmetry between the sheath and magnetosphere plasmas for this event is a factor of $\sim$16. In order to resolve this relatively large asymmetry, macroparticles from the two sides are loaded as separate populations with different numerical weights, selected so that each population is resolved with $\sim150$ particles per species per cell. Besides ensuring reasonable resolution of the low-density plasma, modeling the sheath and magnetosphere plasmas as separate populations also allows the mixing of plasma over time to be diagnosed and quantified \citep{daughton:2014}.

The initial conditions include a drifting Harris sheet population superposed on an asymmetric Maxwellian background \citep{roytershteyn:2012}. For each quantity $Q$ with asymptotic upstream values $Q_{0}$ and $Q_{1}$ on the sheath and magnetosphere sides, we define the following 1D profile depending on coordinate $z$ (note that $x$ and $z$ are reversed compared to typical GSM coordinates at the magnetopause):
\begin{linenomath}
\begin{equation}
F(Q_{0},Q_{1}) = (1/2)[(Q_{1}+Q_{0}) + (Q_{1}-Q_{0})\tanh(z/\lambda)]
\end{equation}
\end{linenomath}
with $\lambda=d_{i0}$ ($d_{i0}$ is the ion inertial length based on sheath density $n_{0}$). The initial reconnecting magnetic field component is $B_x = F(B_{0},B_{1})$ with $B_{1}/B_{0}\sim-1.7$, and the uniform guide field is $B_y/B_{0}\sim0.1$. The temperature profile for each species is taken as $T_s = F(T_{s0},T_{s1})$ with $T_{e1}/T_{e0}\sim3.4$ and $T_{i1}/T_{i0}\sim11.4$ The density profile is then chosen to satisfy hydrodynamic force balance:
\begin{linenomath}
\begin{equation}
n = n_H\textnormal{sech}^2(z/\lambda) + \frac{F(n_{0}T_{0},n_{1}T_{1})}{F(T_{0},T_{1})}
\end{equation}
\end{linenomath}
where the density of the current-carrying Harris population $n_H = \mu_0(|B_1|+|B_0|)^2/8T_H$ with temperature $T_H$ taken as the higher magnetosphere temperature of each species. The density ratio is $n_1/n_0\sim0.062$. The computational domain for the 3D run is $L_x\times L_y\times L_z = 4096\times1024\times2048$ cells $=40d_{i0}\times10d_{i0}\times20d_{i0}$, requiring a total of $\sim2.6$ trillion numerical particles. The $z$ boundaries are conducting and reflect particles, while the domain is periodic in $x$ and $y$, which limits how long the simulation may be run before boundary effects become important \citep{daughton:2006}. An initial magnetic perturbation seeds reconnection with a single dominant X-line. The corresponding 2D run is identical except that it is reduced to a single cell in the $y$ direction. Other parameters are a reduced ion-to-electron mass ratio of $m_i/m_e=100$ and frequency ratio $\omega_{pe0}/\omega_{ce0}=1.5$ (based on sheath conditions). In the following, simulation results are plotted at time $t*\omega_{ci0} = 50$, when reconnection is quasi-steady. The features of our simulation are similar to those of \citet{price:2016}, evidence that the fully developed state of the reconnecting layer depends mainly on the asymptotic boundary conditions and not on the details of the initial equilibrium \citep{pritchett:2009,roytershteyn:2012}.

\begin{figure}
\includegraphics[width = 8.0cm]{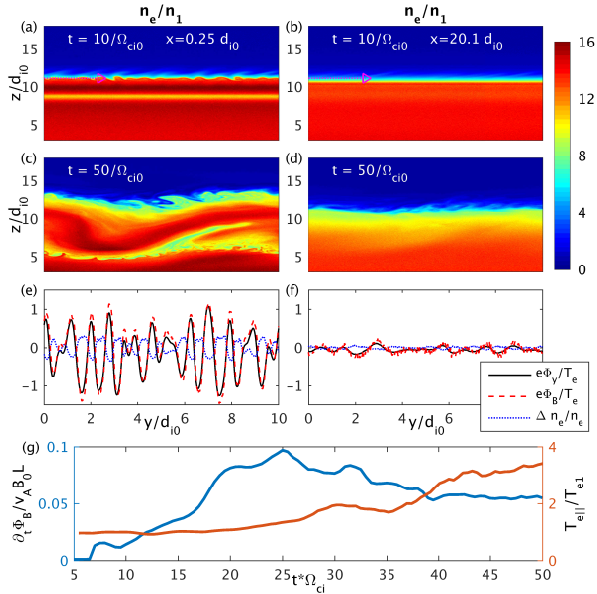}
\caption{Contours of density in the $y$ (direction of current flow) and $z$ (across the current layer) plane at $t\Omega_{ci}=10$ at (a) $x=0.25d_{i0}$ and (b) $x=20.1 d_{i0}$ (X-line). (c-d) Same as (a-b) at later time $t\Omega_{ci}=50$. (e-f) Profiles along the $y$ direction from the horizontal cut at time  $t\Omega_{ci}=10$  indicated in (a-b), where $\beta_e\sim0.25$. The density fluctuations result from flute-like rippling. The potential $\Phi_y =-\int E_y dy$, and $\Phi_B$ is defined by \citet{norgren:2012}. (g) The normalized reconnection rate (blue), and $T_{e||}$ (red) averaged over the $y$ direction and the segment $z=11$---12$d_{i0}$  at $x=20.1d_{i0}$ (X-line). \label{fig:lhdi}}
\end{figure}

\section{Fluctuations, Mixing, and Heating}

The strong density gradient across the magnetopause in this event is susceptible to drift instabilities \citep{price:2016}, which include out-of-plane variations and are absent in 2D models. Based on the typical range of wave vectors with ${\bf{k\cdot B}}\sim0$ and $k_\perp\rho_e\sim1$, the fluctuations and resulting turbulence are associated with the electrostatic lower-hybrid drift instability (LHDI) \citep{daughton:2003}. At early time in the simulation [Figs.~\ref{fig:lhdi}(a) and (c)], a coherent mode that propagates in the electron drift direction is dominant in the simulation with $k_\perp\rho_e\sim0.5$ and $\omega^2\sim\omega_{LH}^2=\omega_{pi}^2/(1+4\pi nm_ec^2/B^2)$, in agreement with a local electrostatic LHDI dispersion relation \citep{krall:1971}. The density profile in the $y-z$ plane is plotted in Fig.~\ref{fig:lhdi} at (a) $x=0.25d_{i0}$ and (b) $x=20.1d_{i0}$ (through X-line), and the variations about the mean density along the indicated horizontal cuts are plotted in Figs.~\ref{fig:lhdi}(e-f). The density is modulated along the magnetopause by the rippling of the density gradient layer.

\begin{figure}
\includegraphics[width = 8.0cm]{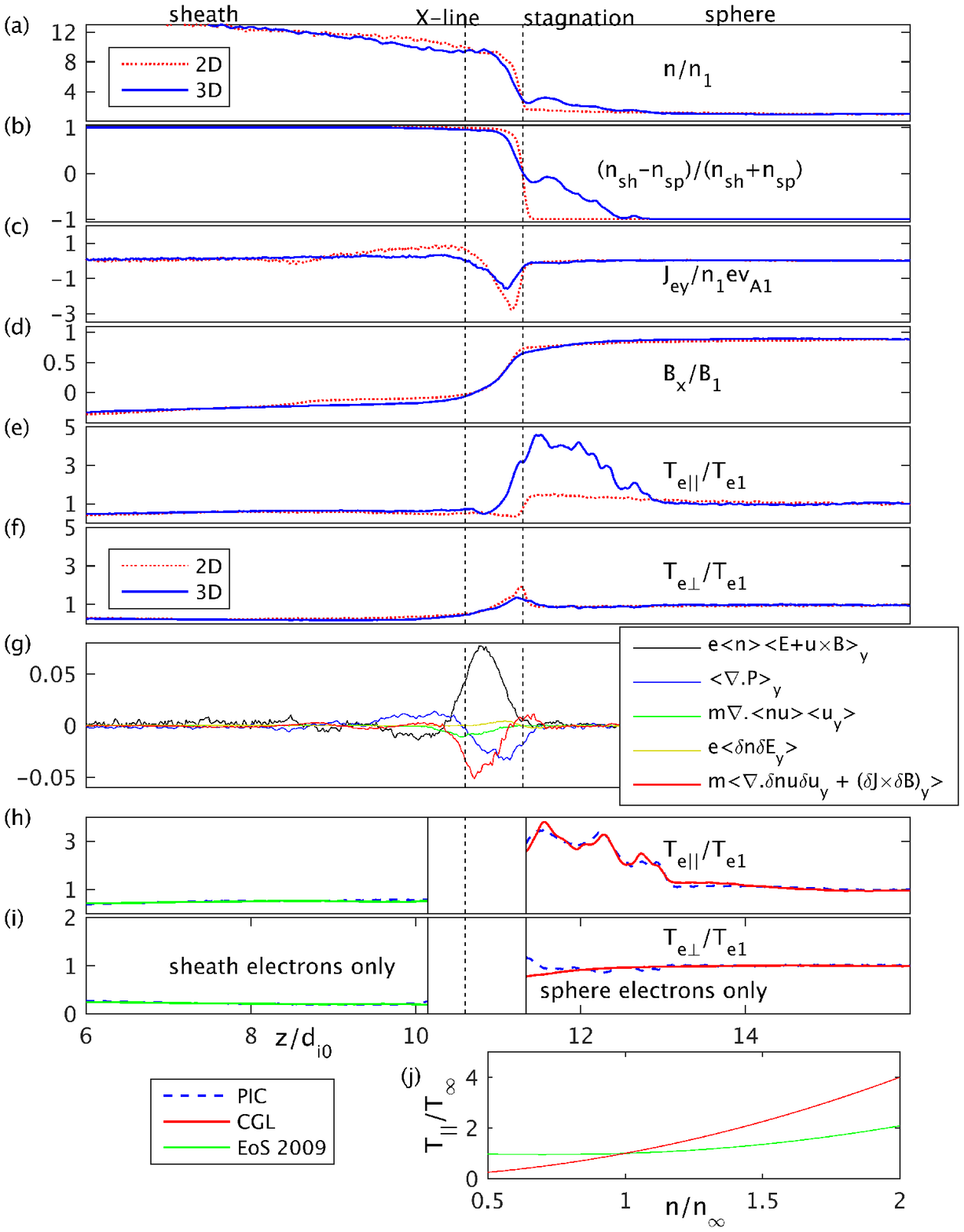}
\caption{Profiles through the X-line at $x=20.1d_{i0}$ across the reconnection layer of (a) the density, (b) the mix diagnostic $M$, (c) the out-of-plane current, (d) the reconnecting magnetic field component, (e) the parallel electron temperature, and (f) the perpendicular electron temperature. (g) Terms in the averaged Ohm's law, where $<Q>$ is a spatial $y$ average of quantity $Q$ and $\delta Q = Q-<Q>$. (h-i) The inflowing sheath electrons display the temperature scalings of \citet{le:2009}, and the magnetosphere electrons follow CGL \citep{chew:1956} scalings. (j) The CGL and trapping equations of state (EoS 2009) for $T_{||}$ as functions of density for a fixed magnetic field strength $B/B_\infty=1$ (subscript $\infty$ refers to upstream conditions). \label{fig:cuts}}
\end{figure}

The LHDI in this low-$\beta$ region is predominantly electrostatic, and the associated electric potential is $\Phi_y=-\int E_y dy$ (although because of the weak guide field, $\bf k$ is not strictly in the $y$ direction). The waves transiently reach a level with $e\Phi_y\sim T_e$ [Fig.~\ref{fig:lhdi}(c)], and at later time $e\Phi_y\sim0.1$---$0.2T_e$ similar to magnetospheric observations \citep{norgren:2012,graham:2016}. Assuming the reconnection electric field is $E_r\sim0.1 B_xv_A/c$, the ratio of the wave electric field $E_w$ to $E_r$ will be
\begin{linenomath}
\begin{equation}
\frac{E_w}{E_r}\sim \sqrt{\beta_e \frac{m_i}{m_e}},
\end{equation}
\end{linenomath}
where we assume $E_w\sim k_\perp \Phi_y \sim \Phi_y/\rho_e$ and $e\Phi_y/T_e\sim0.1$, and $\beta_e$ is the ratio of electron fluid to magnetic pressure. For our initial conditions, $\beta_e$ has asymptotic values of 0.24 (sheath) and 0.017 (magnetosphere). Under typical conditions with $\beta_e>m_e/m_i$, the fluctuating electric field may therefore be significantly larger than the reconnection field and strongly influence the electrons. The contributions of fluctuations to the y-averaged electron momentum balance equation \citep{roytershteyn:2012,price:2016} are plotted in Fig.~\ref{fig:cuts}(g) along a cut across the X-line. We find in agreement with \citet{price:2016} that anomalous viscosity $\propto<\delta J \times \delta B>$ is comparable to the inertial term near the X-line and to the pressure term near the stagnation region. Anomalous resistivity $\propto<\delta n\delta E_y>$ is large during a transient early phase, but it becomes small as the initially steep gradients relax.  This later state is likely more representative of actual conditions in space.

While the lower-hybrid fluctuations here are mainly electrostatic in the edge region, the ${\bf{E\times B}}$ electron currents in the wave field generate a magnetic perturbation of the local equilibrium (reconnecting) field component. \citet{norgren:2012} developed an analysis method based on the following potential inferred from magnetic fluctuations: $(e\Phi_B/T_e) = (2/\beta_e) (\Delta B_x/B_x)$. This potential $\Phi_B$ is compared to the electric potential $\Phi_y$ in Fig.~\ref{fig:lhdi}(c), and they indeed agree well. Note that the magnetic field fluctuation is in phase with the wave electrostatic potential (and thus out of phase with the wave electric field), and $\Delta B_x/B_x$ will be small when $e\Phi_y/T_e < 1$ and $\beta_e<1$. 

At later times, turbulence develops from the fluctuations. The lower-hybrid drift waves appear to couple to electron velocity shear-driven modes \citep{romero:1992}, leading to the development of vortices [Fig.~\ref{fig:lhdi}(b)]. The turning over of vortices may also twist magnetospheric field lines into flux ropes that undergo local reconnection \citep{ergun:2016}, but the specific details of this hypothesis remain to be investigated in the present run. In addition, we observe in Fig.~\ref{fig:lhdi}(b) a kink with the longest wavelength accessible in this system, which based on its wave number of $k_y\sqrt{\rho_i\rho_e}\sim0.6$ could be the electromagnetic LHDI that develops in the high-$\beta$ center of reconnecting current sheets \citep{daughton:2003,roytershteyn:2012}. The kinking of the current sheet is largest at the magnetic O-line within the exhaust, and it is much weaker near the X-line.

\begin{figure*}
\includegraphics[width = 16.0cm]{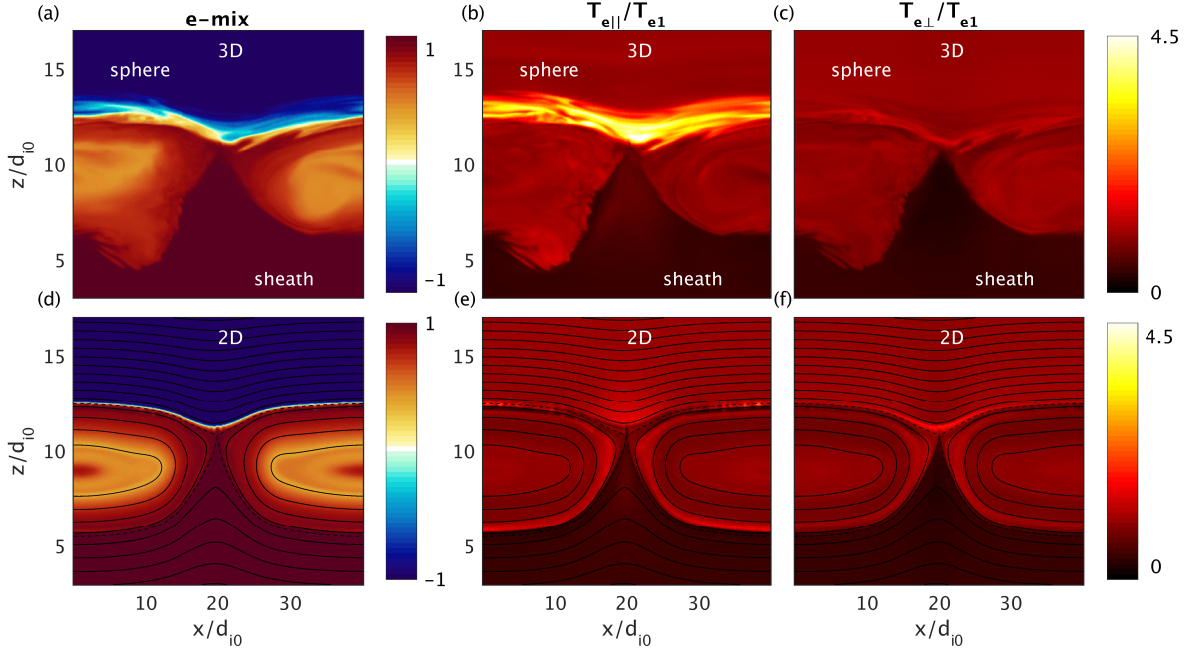}
\caption{Profiles from PIC simulations of asymmetric reconnection. The electron mixing measure in (a) and (d) is defined as $M=(n_{sh}-n_{sp})/(n_{sh}+n_{sp})$, where $n_{sh}$ and $n_{sp}$ are the densities of electrons that were initially on the sheath and magnetosphere sides of the domain. The electron parallel temperature $T_{e||}$ is elevated within the mix layer in the (b) 3D run while it remains lower in the (e) 2D run. The perpendicular electron temperature is moderate in both (c) 3D and (f) 2D. In-plane magnetic flux surfaces are drawn for the 2D simulation. \label{fig:mix}}
\end{figure*}

Although LHDI under typical magnetospheric conditions does not significantly alter the gross reconnection rate or dynamics \citep{zeiler:2002,roytershteyn:2012,pritchett:2013}, the fluctuations may enhance the transport of plasma across the magnetic field. The electron transport is quantified using a mixing diagnostic plotted in Fig.~\ref{fig:mix} from the (a) 3D and (d) 2D simulations, defined as $M=(n_{sh}-n_{sp})/(n_{sh}+n_{sp})$ \citep{daughton:2014}.  The quantities $n_{sh}$ and $n_{sp}$ are the local densities of electrons that originated on the sheath and magnetosphere sides of the initial current layer. Visible in the 3D data [Fig.~\ref{fig:mix}(a)] is a layer of width $\sim $1---2$d_{i0}$ bordering the magnetosphere-side separatrix where the electron populations are well-mixed, with roughly equal electron densities from each side of the magnetopause. In 2D [Fig.~\ref{fig:mix}(d)], on the other hand, the mixing of sheath electrons into the magnetosphere is minimal. In 2D systems, conservation of out-of-plane canonical momentum ties the electron particle orbits to flux surfaces, and particle mixing is therefore limited to a scale on the order of the electron gyroradius (based on the in-plane magnetic field) across the magnetic separatrix.

LHDI initially heats electrons in the perpendicular direction \citep{daughton:2004}, and in large systems it can potentially generate superthermal electrons in the parallel direction \citep{cairns:2005}. In this simulation, the electron energization is mainly parallel and is linked to enhanced transport in 3D, particularly after the reconnection rate peaks [see Fig.~\ref{fig:lhdi}(g)]. The parallel electron temperature $T_{e\parallel}$ is plotted in Fig.~\ref{fig:mix} from the (b) 3D run and the (e) 2D run. In the 3D simulation, the electron parallel temperature is enhanced within the mixing layer, and it reaches peak values of $T_{e\parallel}/T_{e\perp}\sim$ 4---5. One of the key observations by MMS during the 16 October 2015 reconnection event was an increased parallel electron temperature near the diffusion region \citep{burch:2016}, with peak anisotropy in the observations of $T_{e\parallel}/T_{e1}\sim$ 3. Note that the parallel heating is significantly weaker in the 2D run with a peak of $T_{e\parallel}/T_{e\perp}\sim1.6$. In both 2D and 3D, the perpendicular electron heating is comparatively mild [Figs.~\ref{fig:mix}(c) and (f)]. 

The differences between 2D and 3D mixing and heating are also illustrated in Fig.~\ref{fig:cuts}, which displays profiles from both the 2D and the 3D runs along cuts through the approximate X-line. Within the region of strong LHDI fluctuations at $z\sim11$---$13d_{di0}$ in the 3D run, there is an enhancement of the total density [Fig.~\ref{fig:cuts}(a)]. This denser plasma on the magnetosphere side coincides with the region of electron mixing [Fig.~\ref{fig:cuts}(b)], which is shifted to the magnetosphere side of both the approximate X-line (indicated by the vertical dashed lines) and the peak of the out-of-plane current [Fig.~\ref{fig:cuts}(c)] or equivalently the magnetic shear layer [Fig.~\ref{fig:cuts}(d)]. The increased electron parallel temperature that is evident in Fig.~\ref{fig:mix}(b) is also plotted in Fig.~\ref{fig:cuts}(e). As noted above, the perpendicular electron temperature [Fig.~\ref{fig:cuts}(f)] does not display evidence of strong heating in either 2D or 3D.

\section{Particle and Fluid Pictures}

\begin{figure}
\includegraphics[width = 7.5cm]{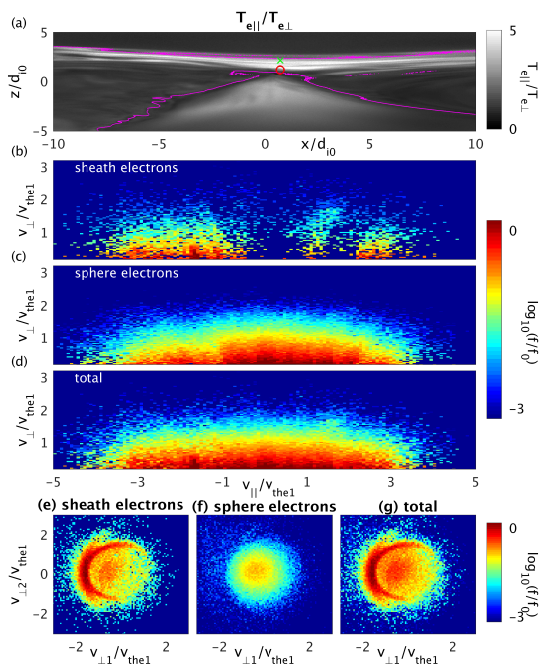}
\caption{(a) The electron temperature anisotropy $T_{e\parallel}/T_{e\perp}$ near the X-line in a cut of the 3D PIC simulation. The magenta contours mark electron mix fractions of 0.1. The distribution in parallel ($v_{\parallel}$) and perpendicular ($v_\perp$) velocity space of electrons originating from the (b) sheath and (c) magnetosphere at the green $\times$ in (a), where $T_{e\parallel}/T_{e\perp}\sim4$. The total combined electron distribution is plotted in (d). (e-g) Similar plots for the point marked by a red $\circ$ in (a) showing perpendicular crescents in $v_{\perp 1}$---$v_{\perp 2}$ space.\label{fig:fe}}
\end{figure}

We consider the energization physics both by examining the electron velocity distributions and by considering a fluid description. Electron velocity distributions in $v_{\parallel}-v_\perp$ space are plotted in Fig.~\ref{fig:fe} from particles in a box of side $0.2 d_{e0}$ centered at the $\times$ plotted in Fig.~\ref{fig:fe}(a). This point, at which $T_{e\parallel}/T_{e\perp}\sim4$, is deeper into the magnetosphere than the typical location of crescent-shaped electron distributions \citep{burch:2016} [see Figs.~\ref{fig:fe}(e-g)], which persist in 3D \citep{price:2016} and are typically found within an in-plane gyroradius of the magnetospheric separatrix \citep{egedal:2016}. 

Separate velocity distributions are plotted in Fig.~\ref{fig:fe} for electrons originating in (b) the sheath and (c) the magnetosphere, along with (d) the total electron distribution. The sheath electrons that have penetrated into the magnetosphere in Fig.~\ref{fig:fe}(b) display a discontinuous distribution as a function of parallel velocity, with gaps for certain ranges of $v_{\parallel}$, suggesting that the plasma conditions within this flux tube vary faster than the transit time of the electrons across the reconnection region. Similar discontinuities are also visible in the magnetosphere electron distribution in Fig.~\ref{fig:fe}(c), where, for example, the phase space density is reduced below $v_\parallel\sim-v_{the1}$. Interestingly, the combined total distribution in Fig.~\ref{fig:fe}(e) is relatively smooth. It resembles the elongated trapped distributions typical of laminar 2D current sheets \citep{egedal:2013pop} and observed near magnetopause reconnection sites \citep{graham:2014,graham:2016}.

In this system, electron orbits very near the X-line are chaotic \citep{le:2013,egedal:2016}. In the inflows, however, the electrons undergo regular magnetized motion with well-conserved adiabatic invariant $\mu$, and those electron populations that remain on one side of the separatrix follow relatively simple equations of state. In Figs.~\ref{fig:cuts}(h) and (i), each electron population (sheath and magnetosphere) is treated as a separate fluid with its own density and temperature moments. The sheath electron density $n_{sh}$ and temperature $T_{e,sh}$ within the sheath follow equations of state for electrons that are adiabatically trapped by a localized parallel electric field \citep{le:2009,egedal:2013pop}. The highest parallel temperatures are predicted in regions of weak magnetic field and enhanced ion density. A weaker magnetic field occurs when magnetic flux tubes expand, driving a rarefaction of the electron density. Meanwhile, the quasi-neutrality constraint requires the electron and ion densities to closely match, and this is accomplished by parallel compression (and associated heating). As plotted on the left-hand sides of Figs.~\ref{fig:cuts}(h) and (i), the trapping leads to moderate parallel heating and perpendicular cooling in the sheath inflow with a peak anisotropy of $T_{esh||}/T_{esh\perp}\sim3$, similar in 2D and 3D. In the 2D simulation, the electron heating and temperature anisotropy are weaker on the low-$\beta$ magnetosphere side, as opposed to previous simulation studies. This is because the in-plane flow stagnation point is well-separated from the X-line \citep{cassak:2007} for these parameters, and as a result few magnetosphere electrons enter regions of enhanced density or weak magnetic field that are conducive to strong heating. 

The heating within the magnetosphere inflow is substantially enhanced in 3D, in part because particle transport alters the plasma profiles [as in Fig.~\ref{fig:cuts}] and magnetosphere electrons enter regions of increased ion density. The heating, however, is stronger than predicted by the adiabatic trapping model, and the magnetosphere electrons (within the magnetosphere inflow) more closely obey CGL scalings \citep{chew:1956}, $T_{e,sp||}\propto n_{sp}^2/B^2$ and $T_{e,sp\perp}\propto B$, as plotted on the right-hand sides of Figs.~\ref{fig:cuts}(h) and (i). Similar scalings hold within the mixing layer downstream of the X-line (not plotted). The CGL model assumes only that the electrons are well-magnetized ($\mu$ is invariant) and the parallel thermal heat flux is negligible. In the deeply-trapped regime, the trapping model of \citet{le:2009} reduces to the CGL scalings, although the CGL parallel temperature is asymptotically a factor of $\sim2$ greater [see Fig.~\ref{fig:cuts}(j)]. The trapping model likely fails because it assumes that trapped electron bounce times are shorter than any other relevant time scale. The ratio of the electron bounce frequency to the lower hybrid frequency, however, is $\sim\omega_{be}/\omega_{LH}\sim 2\pi\sqrt{\beta_e}/(L_\parallel/d_i)$, where a typical parallel length scale $L_\parallel$ is a few $d_i$. Thus, in low- or moderate-$\beta$ magnetospheric plasmas, the finite transit time of electrons must be taken into account for lower-hybrid frequency range fluctuations. In fact, a similar CGL scaling held in a simulation of magnetic island merging \citep{le:2012} within large-amplitude ion cyclotron waves with high parallel phase speeds. This suggests that when the parallel dynamics of fluctuations (with typical frequencies below $\omega_{ce}$) are fast enough, the electron heat flux associated with parallel streaming will become unimportant, and the electron fluid will obey CGL scalings to a good approximation.

\section{Discussion and Summary}

We compared the electron mixing and heating between 2D and 3D kinetic simulations of asymmetric reconnection with plasma parameters matching those of an event observed by the MMS spacecraft \citep{burch:2016}. The primary differences between 2D and 3D resulted from the development of lower-hybrid range fluctuations \citep{daughton:2003}. While the electrostatic lower-hybrid drift waves do not strongly influence the overall reconnection rate \citep{roytershteyn:2012}, they enhance cross-field plasma transport from the sheath into the magnetosphere.

For these plasma parameters, the magnetosphere heating is relatively weak in 2D, as opposed to previous results for asymmetric reconnection \citep{egedal:2011pop}. The difference likely stems from the relatively extreme asymmetry of the upstream plasma conditions, which increases the separation between the X-line and flow stagnation points \citep{cassak:2007}. In 3D, the electron parallel heating is strongest on the magnetosphere side. This results from two mechanisms. First, lower-hydrid induced transport alters the mean plasma profiles, such that magnetosphere electrons move through regions of increased density and weaker magnetic field, which both tend to increase the level of electron parallel heating and temperature anisotropy \citep{le:2009,egedal:2013pop}. And second, the fast parallel dynamics of the lower-hybrid fluctuations leads the electrons to more closely follow CGL temperature scalings, which predict even stronger parallel heating than the adiabatic trapping model [see Fig.~\ref{fig:cuts}(j)]. The CGL scalings may be pertinent to a variety of situations where fluctuations have typical parallel variations faster than electron orbit times, although in practice they would be difficult to apply to observational data when populations of electrons from different sources mix. It is worth noting finally that the transport and heating associated with lower-hybrid fluctuations do not depend strongly on the presence or rate of magnetic reconnection, and they could continue to be important throughout the magnetopause boundary layer even in the absence of reconnection \citep{treumann:1991}.


%
%
%
%
%
%
%

\begin{acknowledgments}
A.L. received support from the LDRD office at LANL and acknowledges NASA grant NNX14AL38G.  W.D.'s work was supported by NASA grant NNH13AW51L. L.-J.C. received support from DOE grant DESC0016278; NSF grants AGS-1202537, AGS-1543598 and AGS-1552142; and the NASA Magnetospheric Multi-scale mission. J.E. acknowledges support through NSF GEM award 1405166 and NASA grant NNX14AC68G. Simulations were performed at LANL on the Trinity machine and with Institutional Computing resources. Archived simulation data are available upon request to the authors.
\end{acknowledgments}

\end{article}
%
%
%
%
%
%
%
%


\end{document}